\documentstyle[12pt]{article}
\def\alt{\hbox{\raise.5ex\hbox{$<$}
\kern-1.1em\lower.5ex\hbox{$\sim$}}}
\def\agt{\hbox{\raise.5ex\hbox{$>$}
\kern-1.1em\lower.5ex\hbox{$\sim$}}}

\def\up#1{\raise 1ex\hbox{\sevenrm#1}}

\begin{document}

\title{\hglue 9cm {\normalsize IHES/P/96/40 \\}
\hglue 9cm {\normalsize IASSNS-HEP-96/62}
\vglue 1cm
The Oklo bound on the time variation of 
the fine-structure constant revisited}
\author{Thibault Damour$^{a,b,c}$, Freeman Dyson$^{b}$\\
$^{a}$ Institut des Hautes Etudes Scientifiques,\\ 
F-91440 Bures-sur-Yvette, France\\
$^{b}$ School of Natural Sciences, Institute for Advanced Study,\\ 
Olden Lane, Princeton, NJ08540, U.S.A.\\
$^{c}$ DARC, Observatoire de Paris--CNRS,\\
 F-92195 Meudon, France.}
\date{June 27, 1996}
\maketitle

\begin{abstract}
It has been pointed out by Shlyakhter that data from the natural
fission reactors which operated about two billion years ago at
Oklo (Gabon) had the potential of providing an extremely tight bound
on the variability of the fine-structure constant $\alpha$. We revisit
the derivation of such a bound by: (i) reanalyzing a large selection
of published rare-earth data from Oklo, (ii) critically taking
into account the very large uncertainty of the temperature at which
the reactors operated, and (iii) connecting in a new way (using isotope
shift measurements) the Oklo-derived constraint on a possible shift of thermal
neutron-capture resonances with a bound on the time variation of $\alpha$. Our
final ($95\%$ C.L.) results are:  $-0.9 \times 10^{-7} <(\alpha^{\rm Oklo} -
\alpha^{\rm now})/\alpha <1.2\times 10^{-7}$ and $-6.7 \times 10^{-17}
{\rm yr}^{-1} < {\dot \alpha}^{\rm averaged}/\alpha<5.0\times10^{-17} {\rm
yr}^{-1}$. 
\end{abstract}

\section{Introduction} \label{sec:1}

Since Dirac \cite{Di37} first suggested it as a possibility, the
time variation of the fundamental constants has remained a subject
of fascination which motivated numerous theoretical and experimental
researches. For general discussions and references to the literature
see, e.g., \cite{Dy72,Dy78,SV90,VP95}. Superstring theories have
renewed the motivation for a variation of the ``constants'' by
suggesting that most of the dimensionless coupling constants of
physics, such as the fine structure constant $\alpha= 1/137.0359895(61)$,
are functions of the vacuum expectation values of some scalar fields
(see, e.g., \cite{GSW}). Recently, a mechanism for fixing the vacuum
expectation values of such {\it massless} stringy scalar fields
(dilaton or moduli) has been proposed \cite{DP94}. This mechanism predicts
that the time variation of the coupling constants, at the present cosmological
epoch, should be much smaller than the Hubble time scale, but maybe not
unmeasurably so. In this model, the time variations of all the coupling
constants are correlated, and the ones of most observational significance are
the fine structure constant $\alpha$ and the gravitational coupling
constant $G$. In the present paper, we revisit the current best bounds on the
variation of $\alpha$.

One of the early ideas for setting a bound on the variation
of $\alpha$ was to consider the fine-structure splittings
in astronomical spectra \cite{savedoff}. With this method, Bahcall and Schmidt
\cite{BS67} concluded that $\alpha$ had varied by at most a fraction
$3\times10^{-3}$ of itself during the last $2\times10^9$ years. A recent update
of this method has given the result $\triangle\alpha/\alpha = (0.2 \pm 0.7)
\times10^{-4}$ at redshifts $2.8\leq z \leq 3.1$, i.e.  the bound $|\dot
\alpha/\alpha| < 1.6\times 10^{-14}{\rm yr}^{-1}$ $(2\sigma \, {\rm
level})$ on the time derivative of $\alpha$ averaged over the last $\sim
10^{10} {\rm yr}$ \cite{VPI96}. See also Ref. \cite{cowie} which obtains $-4.6
\times 10^{-14} {\rm yr}^{-1} < \dot{\alpha} / \alpha < 4.2 \times 10^{-14}
{\rm yr}^{-1}$ from fine-structure splittings, and, denoting $x \equiv \alpha^2
g_p (m_e / m_p)$, $-2.2 \times 10^{-15} {\rm yr}^{-1} < \dot x / x < 4.2 \times
10^{-15} {\rm yr}^{-1}$ by comparing redshifts obtained from hyperfine (21cm)
and optical data.

One of us obtained the upper limit $|\dot
\alpha/\alpha|
< 5\times10^{-15}{\rm yr}^{-1}$ from an analysis of the
abundance ratios of Rhenium and Osmium isotopes
in iron meteorites and molybdenite ores \cite{Dy72}. The most recent direct
laboratory test of the variation of $\alpha$ has obtained
$|\dot \alpha/\alpha| < 3.7\times10^{-14}{\rm yr}^{-1}$ by comparing
hyperfine transitions in Hydrogen and Mercury atoms \cite{PTM95}. For more
references on the variation of constants see \cite{Dy72,Dy78,SV90,VP95}.

On the other hand, the much more stringent bound
$|\dot\alpha/\alpha| < 10^{-17}{\rm yr}^{-1}$ has been claimed by
Shlyakhter \cite{S76a,S76b,S83} to be derivable (at the three standard
deviations level) from an analysis of data from the Oklo phenomenon.
The Oklo phenomenon denotes a natural fission reactor
(moderated by water) that operated about two billion years ago in the
ore body of the Oklo uranium mine in Gabon, West Africa. This phenomenon
was discovered by the French Commissariat \`a l'Energie Atomique (CEA)
in 1972. The results of a thorough, multi-disciplinary investigation
of this phenomenon have been presented in two conference proceedings
\cite{Oklo1,Oklo2}.  See also \cite{M76} and \cite{P77} for summaries of
the first phase of investigation.

In view of the importance of Shlyakhter's claim, of
the lack of publication of a detailed analysis \footnote{The very
brief account published in Nature \cite{S76a} omits most of the analysis
that is presented in the two preprints \cite{S76b,S83}.}, and of our
dissatisfaction with some important aspects of the analysis presented
in two preprints \cite{S76b,S83}, we decided to revisit the Oklo bound on
$\alpha$.
The main conclusions of our work are the following: (i) we confirm the
basic claim of Shlyakhter that the Oklo data is an extremely sensitive
probe of the time variation of $\alpha$; (ii) after taking into account various
sources of uncertainty (notably temperature effects) in the analysis
of data, and connecting in a improved way the raw results of this
analysis to a possible variation of $\alpha$, we derive what we think
is a secure (95\% C.L.) bound on the change of $\alpha$:
\begin{equation}
-0.9 \times 10^{-7} < \frac{\alpha^{\rm Oklo}-\alpha^{\rm now}}{\alpha} <
1.2\times10^{-7} \, .
       \label{eq:1.1}
\end{equation}
In terms of an averaged rate of variation, this reads
\begin{equation}
-6.7 \times 10^{-17} {\rm yr}^{-1} < 
\frac{\dot\alpha}{\alpha} < 5.0\times10^{-17}{\rm yr}^{-1} \, .
\label{eq:1.2} 
\end{equation}

\section{Extracting the neutron capture cross section of Samarium 149
from Oklo data}   \label{sec:2}

The proof of the past existence of a spontaneous chain reaction in the
Oklo ore consists essentially of:  (i) a substantial depletion of the
Uranium isotopic ratio \footnote{For typographical convenience, we
indicate atomic mass numbers as right, rather than left, superscripts.}
${\rm U}^{235}/{\rm U}^{238}$ with respect to the current standard value
in terrestrial samples;  and (ii) a correlated peculiar distribution of
some rare-earth isotopes. The rare-earth isotopes are
abundantly produced in the fission of ${\rm U}^{235}$ and the observed
isotopic distribution is beautifully consistent with calculations of the
effect of a strong neutron flux on the fission yields of ${\rm U}^{235}$
(see e.g. \cite{1.371,1.385,1.541,R76,2.441,2.473}).  In particular, the
strong neutron absorbers Sm$^{149}$, Eu$^{151}$, Gd$^{155}$ and
Gd$^{157}$ are found in very small quantities in the central regions of
the Oklo reactors (see, e.g., Fig.~2 of \cite{R76}). These isotopes were
evidently burned up by the large neutron fluence produced by the fission
process. Following Shlyakhter's suggestion \cite{S76b,S83}, we concentrate on the
determination of the neutron capture cross section of Sm$^{149}$: 
Sm$^{149}(n,\gamma)$ Sm$^{150}$.

The evolution of the concentrations of the various Samarium isotopes
(sharing the common atomic number $Z=62$) in the Oklo ore is especially
simple to describe because of the absence of a stable chemical element
with atomic number $Z=61$.  The most stable Promethium nuclide is
Pm$_{61}^{145}$ with a half-life of 17.7 years.  If there had existed,
before the reaction started, some natural concentration of Promethium it
could, via neutron absorption and subsequent $\beta^{-}$ decay, have generated
some Samarium.  In absence of this, the final values of the Sm
concentrations are determined by:  (i) their initial concentrations,
before the reaction;  (ii) the yields from the fissions;  and (iii) the
effect of neutron captures.

Following Refs.~\cite{Oklo1,Oklo2}, one characterizes the neutron absorbing
power of an isotope by the {\it effective} cross section
\begin{equation}
 \hat\sigma \equiv \frac{\int \sigma(E)vn_{E} dE}{v_0\int n_{E}dE}
 \label{eq:2.1}
\end{equation}
where $v$ is the (relative) velocity of incident neutrons, $n_{E}dE$ the
energy distribution of the neutrons, and $v_0$ the fiducial (thermal)
velocity $v_{0}=2200$m/s corresponding to a kinetic energy $E_{0}=0.0253$eV.
The advantage of the definition (\ref{eq:2.1}) is that, in the case of a
``$1/v$ absorber'', $\sigma(E)=C/v$, the effective cross section equals
$\hat\sigma=C/v_{0}=\sigma(E_0)$ {\it independently} of the neutron spectrum.
[To a good appoximation, this is the case for the thermal fission cross
section of ${\rm U}^{235}$]. On the other hand, in the case of nuclides
exhibiting resonances in the thermal region (these are the strong
absorbers, Sm$^{149}$, Eu$^{151}$, Gd$^{155}$, Gd$^{157}$), the value of the
effective cross section (\ref{eq:2.1}) is very sensitive to the neutron
spectrum, especially to its thermal part \footnote{
The spectrum of moderated neutrons in a fission reactor consists of a
Maxwell-Boltzmann thermal distribution up to energies of order a few
times $kT$, followed by a tail $n_{E}dE \propto dE/(vE)$ due to neutrons
still in the process of moderation}.

Associated to the introduction of the effective cross section
(\ref{eq:2.1}), one defines an effective neutron flux $\hat\phi\equiv nv_0$ 
with $n=\int n_{E}dE$, and an effective infinitesimal fluence (integrated
flux) \begin{equation}
   d\tau = \hat\phi dt=nv_{0}dt\ .   \label{eq:2.2}
\end{equation}
With this notation, the general equation describing the evolution of
the total number $N_A$ of nuclides of mass number $A$ (for some fixed
atomic number $Z$) in some sample reads
\begin{equation}
  {d N_A\over d\tau}=y_A N_5 \sigma_{f5} + \sigma_{A-1} N_{A-1}
- \sigma_A N_A \, .   \label{eq:2.3}
\end{equation}

Here, $y_A$ denotes the yield of the element $A$ in the fission of
${\rm U}^{235}$, $N_5$ and $\sigma_{f5}$ are short hands for $N_{235}$ and
the (effective) fission cross section of ${\rm U}^{235}$, and the last two
terms describe the effects of neutron captures within isotopes of some
chemical element $Z$.  For simplicity, we drop the carets over the cross
sections.  The evolution equation (\ref{eq:2.3}) neglects any contribution
$\propto N_{A-1}(Z')$ coming from the $\beta^-$ decay of the neighbouring
chemical element $Z'=Z-1$ after absorption of a neutron.  As we said
above, this approximation applies well to the Samarium case. Eq.~(\ref{eq:2.3})
neglects also the yields due to the fractionally small number of
fissions of ${\rm U}^{238}$ and Pu$^{239}$.  [See, e.g., Ref.~\cite{R76}
which estimates that, in a particular sample, 2.5\% and 3\% of the fissions
were due to ${\rm U}^{238}$ and Pu$^{239}$, respectively].

Samples in the cores of the various Oklo reactors were exposed
to a total effective fluence $\tau=\int d\tau=\int nv_{0}dt$ of the order
of $10^{21}$neutron/cm$^2 = 1$ inverse kilobarn. This means, roughly
speaking, that processes with effective cross sections comparable or larger
than 1 kb have led to a significant number of reactions, while
processes with ${\sigma \ll 1{\rm kb}}$ had a negligible effect. The
former category includes the fission of ${{\rm U}^{235} (\sigma_{f5} \sim
0.6 {\rm kb})}$, the capture of neutrons by ${\rm Nd}^{143} (\sigma_{143} \sim
0.3 {\rm kb})$ and by the strong absorbers (such as ${\rm Sm}^{149}
; \sigma_{149} \agt 70 {\rm kb})$, while the latter category includes neutron
captures by weak absorbers such as ${\rm Sm}^{144}$ and ${\rm Sm}^{148}$ with
cross sections of only a few barns.

This allows one to neglect $\sigma_{144}$ and $\sigma_{148}$ (for $Z=62)$
in Eq.~(\ref{eq:2.3}). Further simplification comes from the fact
that the stable isotopes 144, 146 and 148 of Neodymium prevent the
formation of the long-lived\footnote{The half-life of Sm$^{146}$ is
$1.03\times 10^8$yr and therefore long with respect to the duration of
the Oklo phenomenon.} Sm$^{144}$, Sm$^{146}$ and Sm$^{148}$ as end
points of ($\beta^-$ decay) fission chains. The stable
isotopes\footnote{``Stable'' means, in this context, a half life much larger
than the {\it age} of the Oklo phenomenon. E.g. the half-life of Sm$^{147}$ is
$1.06\times 10^{11}$yr$\gg 2\times 10^9$yr.} obey the simple evolution
equations 
\begin{eqnarray}
 {dN_{144}\over d\tau} &=& 0 \ , \label{eq:2.4a} \\
 {dN_{147}\over d\tau} &=& y_{147} N_5 \sigma_{f5} -\sigma_{147} N_{147}\ ,
   \label{eq:2.4b} \\
 {dN_{148}\over d\tau} &=& \sigma_{147} N_{147} \ , \label{eq:2.4c} \\
 {dN_{149}\over d\tau} &=& y_{149} N_5 \sigma_{f5} -\sigma_{149} N_{149}\ ,
   \label{eq:2.4d} \\
 {dN_{5}\over d\tau} &=& -N_5 \sigma^*_5  \ . \label{eq:2.4e}
\end{eqnarray}
To close the system, we have followed \cite{1.541} and \cite{R76} in describing
the burn up of ${\rm U}^{235}$ by means of a modified absorption cross
section $\sigma^*_5=\sigma_5 (1-C)$, where $\sigma_5$ is the normal
absorption cross section (fission plus capture) and $C$ is a
conversion factor representing the formation of ${\rm U}^{235}$
from the decay of Pu$^{239}$ formed by neutron capture in
${\rm U}^{238}$.

In the approximation where the (effective) cross sections, and the conversion
factor $C$, are constant, the system (\ref{eq:2.4a}--\ref{eq:2.4e}) is easily
solved and gives
\begin{eqnarray}
  N_5 (\tau) &=& N_5 (0) e^{-\sigma^*_5\tau}\ , \label{eq:2.5a}\\
  N_{144} (\tau) &=& N_{144} (0) \ , \label{eq:2.5b}\\
  N_{147} (\tau) + N_{148} (\tau)
    &=& N_{147} (0) + N_{148} (0) + y_{147} \sigma_{f5} N_5(0)
   {1-e^{-\sigma^*_5\tau}\over \sigma^*_5}\ , \label{eq:2.5c}\\
  N_{149} (\tau)
    &=& N_{149} (0) e^{-\sigma_{149}\tau} +y_{149} \sigma_{f5} N_5(0)
   {e^{-\sigma^*_5\tau} -e^{-\sigma_{149}\tau} \over
    \sigma_{149} -\sigma^*_5} . \label{eq:2.5d}
\end{eqnarray}
 Eq.~(\ref{eq:2.5b}) shows that the quantity of Sm$^{144}$ measured
in a sample now is equal to the quantity
of natural Sm$^{144}$ present in the sample before the nuclear
reactions. Assuming that the natural Samarium present in the sample at
the beginning had the normal isotopic ratios (say $n_{144} =3.1\%$,
$n_{147} =15.0\%$, $n_{148}=11.3\%$, $n_{149}=13.8\%$, etc...
\cite{GE}), we can use (\ref{eq:2.5b}) to correct (\ref{eq:2.5c}) for the
initial concentrations in Sm$^{147}$ and Sm$^{148}$. The effect of
$N_{149}(0)$ in Eq.~(\ref{eq:2.5d}) is totally negligible (as is the
last term) because the exponent $\sigma_{149}\tau \gg 1$. We then derive
the intermediate result
\begin{equation}
 {N_{147} (\tau) + N_{148} (\tau) - {n_{147} +n_{148}\over n_{144}}
  N_{144}(\tau) \over N_{149} (\tau)} = {y_{147}\over y_{149}}
  {e^{\sigma_5^* \tau} -1 \over \sigma^*_5} (\sigma_{149}-\sigma^*_5)\ .
  \label{eq:2.6}
\end{equation}

 One can finally obtain an expression for $\sigma_{149}$ in terms of
``measured'' quantities by connecting $\sigma_5^*\tau$ to the observed
ratio between the numbers of ${\rm U}^{235}$ and ${\rm U}^{238}$ atoms
in the Oklo sample. If, following \cite{1.541}, we define
\begin{equation}
 w\equiv {0.00725\over (N_5/N_8)^{\rm Oklo}_{\rm now}} \ , \label{eq:2.7}
\end{equation}
where $0.00725$ is the usual ${\rm U}^{235}/{\rm U}^{238}$ ratio in natural
Uranium now, it is easy to verify that $w=e^{\sigma^*_5\tau}$. Finally, we
get \footnote{Eq.~(\ref{eq:2.8}) is equivalent to equations appearing in
Refs.~\cite{1.541,S76b,S83} apart from the facts that Shlyakhter's
equations contain misprints (e.g. $(w -1)/\ln w$ instead of its inverse
in Eq.~(\ref{eq:2.8})). The fractionally small first contribution on the
right-hand side of (\ref{eq:2.8}) is neglected in the above references.}
\begin{equation}
 \sigma_{149} = {1\over \tau} \left[ \ln w + y {\ln w\over w-1}
   {N_{147}+N_{148}-n N_{144}\over N_{149}} \right]\ , \label{eq:2.8}
\end{equation}
where (using Refs.~\cite{Rider} and \cite{GE})
\begin{eqnarray}
 y &=& {y_{149}\over y_{147}} = {1.080384\over 2.261681} \simeq 0.478\ ,
  \label{eq:2.9} \\
 n &=& {n_{147}+n_{148}\over n_{144}} = {15.0+11.3\over 3.1} \simeq 8.48\ .
  \label{eq:2.10}
\end{eqnarray}
The quantities $N_A$ in Eq.~(\ref{eq:2.8}) denote the present values of
the isotopic concentrations, or, equivalently, the present values of the
isotopic ratios. The isotopic ratios of Samarium have been measured in many
Oklo samples \cite{1.357,1.371,R76,2.433,2.441}. Note that the quantity
which is, at this stage, directly obtainable from observations is the
dimensionless product $\sigma_{149}\tau =\int dt \int \sigma_{149}(E) vn_E
dE.$

Similarly, by considering the fission yields of Neodymium and the
neutron-capture reaction Nd$^{143}\to$Nd$^{144}$ (using, e.g.,
the Nd$^{142}$ content to subtract the contribution from the natural
concentrations present before the reaction), several authors
\cite{1.371,1.385,1.541} have shown how to obtain the dimensionless
product $\sigma_{143}\tau$ (where $\sigma_{143}\equiv \sigma_{(n,\gamma)}
({\rm Nd}^{143})$) in terms of quantities observed in Oklo samples.
Combining these two results, we see that the value two billion years ago of
the ratio $\sigma_{149}\tau / \sigma_{143}\tau = \sigma_{149}/\sigma_{143}$ can
be computed in terms of present Oklo data.

Although it would be conceptually clearer to deal only with the
dimensionless ratio $\sigma_{149}/\sigma_{143}$, we shall follow
previous usage in working with the dimensionful quantity $\sigma_{149}$
obtained by inserting in Eq.~(\ref{eq:2.8}) the value of
the (effective) fluence $\tau$ deduced from Neodymium data by previous
authors. This procedure is justified by the fact that the effective
cross-section $\sigma_{143}$, defined by Eq.~(\ref{eq:2.1}), depends
very little on the neutron spectrum because $\sigma_{143}(E)$ follows
the $1/v$ law over most of the range of interest. Therefore the lack of
knowledge of the temperature of the moderated neutrons is of no
importance (contrary to the case of $\sigma_{149}$) and the effect of
epithermal neutrons is also very small\footnote{In the analysis of Oklo
data, it has been customary to parametrize the contribution of
epithermal neutrons to the spectrum by a parameter called $r$. This parameter
is found to be small, $r\sim 0.15$, and its effect on $\sigma_{143}$ is
only a few percent \cite{2.407}.}. In other words, the extraction of
$\sigma_{143}\tau$ from Oklo data is approximately done by assuming a
fixed, fiducial value for $\sigma_{143}$, say $\sigma_{143}\simeq 325$b,
so that the use of Eq.~(\ref{eq:2.8}) for computing a dimensionful
$\sigma_{149}$ is approximately equivalent to computing the
dimensionless quantity 325$\sigma_{149}/\sigma_{143}.$

The detailed isotopic analysis of Oklo data \cite{Oklo1,Oklo2} has shown
that, generally speaking, the ore composition has changed very little
since the end of the nuclear reactions. This is established by studying
the correlation between the fluence $\tau$ and the Uranium isotopic
ratio $N_5/N_8$, and by showing that it can be explained by neutronics
considerations (see, e.g. \cite{2.433}).
However, in some cases there is evidence for a partial reshuffling of
chemical elements after the end of the reactions. We have examined
these results and selected 16 samples as especially suitable
for extracting a reliable value of $\sigma_{149}$. These samples are all
core samples with high Uranium content, large depletions of ${\rm
U}^{235}$, and high fluences, $\tau\agt  0.7\times 10^{21}$n/cm$^2$. In all
cases, the natural element correction in Eq.~(\ref{eq:2.8}) is small, the
observed Samarium having been produced almost entirely by fission.
The very small content of Sm$^{149}$ (and, when data are available, of
other strong absorbers such as Gd$^{155}$ and Gd$^{157}$) is also a
confirmation of the absence of chemical reshuffling after the reaction. The
data we took come from \cite{1.357,1.541,R76,2.433,2.441}, and
\cite{2.553}\footnote{The sample SC521472 taken from this last reference
was exposed to a smaller fluence than the others. It was included because this
sample has been used to estimate the temperature of the neutrons.}.
The result of calculating $\sigma_{149}$ from these data is exhibited
in Table~I.

The large scatter of the values exhibited in Table~I is compatible
with the strong temperature dependence of $\sigma_{149}$ (see
below). The only exception is the 36kb obtained for the sample
SC39--1387. This value is a clear outlier which, most plausibly, has
been contaminated in some way. Excluding this result, the other 15 results
are all contained in the range
\begin{equation}
 57~{\rm kb} \leq \hat\sigma_{149} \leq 93~{\rm kb}\ . \label{eq:2.11}
\end{equation}
We think that it is conservative to consider the full range
(\ref{eq:2.11}) as a ``2$\sigma$'' (or 95\% C.L.) interval for
$\hat\sigma_{149}$. [For clarity, we reestablish the caret meaning that
we are dealing with an effective cross section, Eq.~(\ref{eq:2.1}).]
Actually, in view of what is known from Oklo, it is very plausible that
the range (\ref{eq:2.11}) is to be attributed to a mixture of
temperature effects and a small amount of post-reaction chemical
reshuffling. For our purpose, we will use the full range (\ref{eq:2.11})
to define a conservative bound on the variation of $\alpha$. For
completeness, and in view of the special use we make of the interval
(\ref{eq:2.11}) we give in Table~II the complete set of data allowing
one (using (\ref{eq:2.8})) to compute $\hat\sigma_{149}$ for the samples
giving the extreme values (\ref{eq:2.11}).

Let us note that the values we obtain for $\hat{\sigma}_{149}$ are
different from the result claimed by Shlyakhter \cite{S76b,S83}, namely
$\hat\sigma_{149}=(55\pm 8){\rm kb}$. As he did not mention the data he
used, we could not trace the origin of this
difference. We note that most of the values in Table~I are compatible
with thermal effects ($\hat\sigma_{149}$ increases from $\sim 70{\rm
kb}$ to $\sim 99{\rm kb}$ when the temperature varies from 20$^\circ$C
to $\sim 400^\circ$C, and then decreases for higher
temperatures\footnote{The fact that we did not find values between 93 and
99kb is probably explained by some post-reaction remobilization. Anyway, the
limit we shall derive on $\alpha$ depends only on the lower bound on
$\hat{\sigma}_{149}$.}).

\section{Bounding a possible shift of the lowest resonance in the capture
cross section of Samarium 149}  \label{sec:3}

Following Shlyakhter's suggestion \cite{S76a,S76b,S83}, we shall
translate the range of ``Oklo'' values of $\hat\sigma_{149}$,
Eq.~(\ref{eq:2.11}), into a bound on the possible shift, between the time
of the Oklo phenomenon and now, of the lowest resonance in the
monoenergetic cross section $\sigma_{149}(E)$. The large values of the thermal
capture cross sections of Sm$^{149}$, Gd$^{155}$ and Gd$^{157}$ are due to the
existence of resonances in the thermal region. In presence of such a resonance,
the monoenergetic capture cross section is well described, in the thermal
region, by the Breit-Wigner formula
\begin{equation}
 \sigma_{(n,\gamma)} (E) = \pi {\hbar^2\over p^2} g {\Gamma_n(E)
\Gamma_\gamma\over (E-E_r)^2 + {1\over 4} \Gamma^2}\ . \label{eq:3.1}
\end{equation}
Here $p$ is the momentum of the neutron, $E=p^2/2m_n$ its kinetic
energy, $g=(2I'+1) (2s+1)^{-1} (2I+1)^{-1}$ a statistical factor depending
upon the spins of the compound nucleus $I'$, of the incident neutron
$s={1\over 2}$, and of the target nucleus $I$, $\Gamma_n(E)$ is a
neutron partial width, $\Gamma_\gamma$ a radiative partial width, and
$\Gamma$ the total width. The neutron partial width $\Gamma_n(E)$ varies
approximately as $E^{1/2}$
\begin{equation}
 \Gamma_n (E) = {2\gamma^2_n\over \hbar} p\ , \label{eq:3.2}
\end{equation}
where $\gamma_n^2$ is a ``reduced partial width'' (see, e.g., \cite{WW58}).
With sufficient approximation, the total width is given by
\begin{equation}
 \Gamma \simeq \Gamma_\gamma + \Gamma_n (E_r)\ . \label{eq:3.3}
\end{equation}

As we shall see explicitly below, the position of the resonance $E_r$ (with
respect to the threshold defined by zero-kinetic-energy incident neutrons)
is extremely sensitive to the value of the fine-structure constant.
By contrast, the other quantities entering the Breit-Wigner formula,
$\gamma^2_n$, $\Gamma_\gamma$, have only a mild (polynomial) dependence
on $\alpha$. Therefore the sensitivity to $\alpha$ of the effective
cross sections $\hat\sigma$, Eq.~(\ref{eq:2.1}), measured with Oklo data
is totally dominated, for strong absorbers, by the dependence of $\sigma (E)$
upon the position of the lowest lying resonance $E_r$. As we said above,
we should, more rigourously, work with dimensionless ratios such as
$\hat\sigma_{149} / \hat\sigma_{143}$. However, the mild absorber Nd$^{143}$
has no resonances in the thermal region. Therefore the $\alpha$-sensitivity
of the ratio $\hat\sigma_{149}/\hat\sigma_{143}$ is completely dominated
by the $\alpha$-sensitivity of $\hat\sigma_{149}$ inherited from the
dependence on $E^{149}_r (\alpha)$.

The main problem is to use the range (\ref{eq:2.11}) of values of
$\hat\sigma_{149}$ to put a limit on a possible shift of the lowest lying
resonance in Sm$^{149}$, 
\begin{equation}
  \Delta \equiv E^{149{\rm (Oklo)}}_r - E_r^{149{\rm (now)}} \ .
\label{eq:3.4} \end{equation}
Previous attempts \cite{S76a,S76b,S83,SV90} at relating
Oklo-deduced ranges of values of $\hat\sigma_{149}$ to $\Delta$ are
unsatisfactory because they did not properly take into account the very
large uncertainty in $\hat\sigma_{149}$ due to poor knowledge
of the neutron temperature in the Oklo reactors. The original analysis
of Shlyakhter assumed a temperature $T\simeq 20^\circ$C (which is much
too low), and the analysis of \cite{SV90} took $T\simeq$~1~000~K, i.e.
$T\simeq 725^\circ$C (which is possible, but on the high side) and assumed
that one could work linearly in the fractional shift $\Delta/E_r$.
We think that one should neither fix the neutron temperature
$T$ (which could have varied over a wide range), nor work linearly
in $\Delta/E_r$ (which could have been larger than unity).

 Several studies, using independent data, have tried to constrain the value
of the temperature in the Oklo reactors \cite{Oklo2}. Mineral phase
assemblages observed within a few meters of the Oklo reactor zones 2 and 5
indicate a minimum temperature in these regions of about 400$^\circ$C, while
relict textures in the reactor zone rock suggest that temperatures
$T\simeq 650-700^\circ$C may have been reached within the reactors
\cite{2.235}. A study of fluid inclusions and petrography of the sandstones
suggest pressures $p\simeq 800-1000$ bar\footnote{It is thought that the
Oklo phenomenon took place while the Uranium deposits were buried $\sim
4$~km deep \cite{2.636}.}
and temperatures ranging between 180$^\circ$C and at least 600$^\circ$C.
On the other hand, the temperature of the water-moderated neutrons
during the fission reactions has been evaluated by a study of the
Lu$^{176}$/Lu$^{175}$ and  Gd$^{156}$/Gd$^{155}$ isotope ratios in several
samples. The values obtained range between 250$\pm$40$^\circ$C and
450$\pm$20$^\circ$C depending upon the sample and the isotope ratio
considered \cite{2.553}. It is to be noted that the concentration of strong
absorbers such as Sm$^{149}$ or Gd$^{155}$ (which are burned very 
efficiently) is determined by the values of the effective cross sections
$\hat\sigma_{149}$ or $\hat\sigma_{155}$  {\it at the end of the fission
phenomenon}. Therefore, we cannot exclude that the temperature to be
used in evaluating $\hat\sigma_{149}$ or $\hat\sigma_{155}$ be on the
low side of the allowed range. 

Summarizing, we consider that the
temperature to be used to determine $\hat\sigma_{149}$ or
$\hat\sigma_{155}$  could be in the full range
\begin{equation}
   180^\circ {\rm C} \leq T \alt 700^\circ {\rm C}\ . \label{eq:3.5}
\end{equation}
Inserting a Maxwell-Boltzmann spectrum \footnote{We do not consider the
effect of epithermal neutrons, which introduce only a rather small
fractional correction (spectrum index of order $r\sim 0.15$
\cite{Oklo1,Oklo2}). This correction is negligible compared to the wide
range we consider.}
\begin{equation}
 {n_E\over n} dE = {2\pi\over (\pi k T)^{3/2}} e^{-{E\over kT}} E^{1/2} dE\
,
 \label{eq:3.6}
\end{equation}
in the definition (\ref{eq:2.1}), with $\sigma (E)$ of the Breit-Wigner
form (\ref{eq:3.1}), we find that the dependence of the effective cross
section of a strong absorber on resonance shift $\Delta$,
Eq.~(\ref{eq:3.4}), and temperature is given by
\begin{equation}
 \hat\sigma (\Delta,T) = {2\pi\over (\pi kT)^{3/2}} \sigma_0 (1+y^2_0)
  \int^\infty_0 {e^{-{E\over kT}} E^{1/2} dE\over 1+y^2 (E,\Delta)}\ .
  \label{eq:3.7}
\end{equation}
Here $\sigma_0$ denotes the ``thermal'' radiative cross section (as observed
now), i.e. the monoenergetic $\sigma_{(n,\gamma)} (E_0)$ evaluated at
$E_0\equiv 0.0253$eV, $y_0$ denotes $2(E_0-E^{\rm now}_r)/\Gamma$, and
$y(E,\Delta) \equiv 2(E-E_r^{\rm now} -\Delta)/\Gamma$. 
The $E^{1/2}$ in the integrand comes from combining several different factors:
a factor $E^{-1/2}$ coming from $\Gamma_n(E)/p^2$ $(1/v$ law), a factor
$E^{1/2}$ coming from the factor $v$ in Eq.~(\ref{eq:2.1}), and the factor
$E^{1/2}$ in the Maxwell spectrum (\ref{eq:3.6}).

In the case of Sm$^{149}$ the numerical values needed to evaluate
(\ref{eq:3.7}) are (from \cite{M84}): $E^{\rm now}_r =0.0973$eV,
$\sigma_0=40.14$kb, and $\Gamma \simeq 0.061$eV. [The latter being
estimated from $\Gamma\simeq \Gamma_\gamma +\Gamma_n (E_r)$ with
$\Gamma_\gamma = 60.5\times 10^{-3}$eV,
$2g \Gamma_n (E_r) = 0.6\times 10^{-3}$eV, with $2g=9/8$ corresponding
to $I'=4$ and $I=7/2$.] The dependence of $\hat\sigma_{149}$ upon the
resonance shift $\Delta$ is shown in Fig.~1 for several temperatures
spanning the range (\ref{eq:3.5}). On the same Figure, we have
indicated the conservative range of values of $\hat\sigma_{149}$,
Eq.~(\ref{eq:2.11}).

The limits on $\Delta$ shown in Fig.~1 are
$-0.12$eV$<\Delta <0.08$eV. The lower limit depends on the minimal allowed
temperature. Given the temperature estimates quoted above, we consider
180$^\circ$C as a firm minimal temperature and therefore $-0.12~{\rm eV}$ as a
firm lower bound. As the upper limit 0.08eV depends on the maximum allowed
temperature which is more uncertain, we have also explored temperatures higher
than 700$^\circ$ C. 
We found that when $\Delta =0.09$eV $\hat\sigma$ never exceeds
57kb even if $T$ is allowed to take values much larger than 700$^\circ$C.
[$\hat\sigma (0.09, T)$ reaches a maximum $<57$kb somewhere around $T\sim
1000^\circ$C.] Therefore to be conservative, we take $\Delta < 0.09$eV as firm
upper limit. We conclude that the Oklo Samarium data constrain a possible
resonance shift to be in the range   \begin{equation}
  -0.12~{\rm eV} < \Delta < 0.09~{\rm eV}\ . \label{eq:3.8}
\end{equation}
For comparison, let us mention that Refs~\cite{S76a,S76b,S83} estimate
a $2\sigma$ range $|\Delta| <0.02$eV from the Samarium data alone, and a
$3\sigma$ range $|\Delta| <0.05$eV from combining Samarium and Europium data.

We tried to make use of Oklo Gadolinium data to restrict further the
Samarium-derived range (\ref{eq:3.8}). A priori, one could think of making use
of both Gd$^{155}$ and Gd$^{157}$ which are strong absorbers of neutrons. In
fact, Gd$^{157}$ is such a strong absorber that its final concentration
$N_{157} \propto y_{157}/\hat\sigma_{157}$ (generalizing
Eq.~(\ref{eq:2.5d})) is too small to be measured reliably. The case of
Gd$^{155}$ is more favorable, its effective cross section being comparable
to that of Sm$^{149}$. However, its fission yield is much smaller ($y_{155}
=0.032\%$ instead of $y_{149}=1.08\%$). The absolute concentration of
all isotopes of Gadolinium is about ten times smaller than that of
Samarium (see, e.g., \cite{R76}). This implies that Gadolinium data are
much more prone to various contaminations (natural element contamination
due to a post-reaction remobilization, and uncertainties in the isotopic
analysis measurements). To make a meaningful analysis of Gadolinium
data, one should probably restrict oneself to samples that were exposed
to rather mild fluences. Such samples are SC361901 and SC521472 which
have been studied in detail in \cite{2.553}. The effective cross
sections $\hat\sigma$(Gd$^{155}$) obtained in the latter reference are
$\hat\sigma_{155}=(42.0\pm 0.5)$kb in SC~361901, and
$\hat\sigma_{155}=(32.5\pm 0.5)$kb in SC~521472. These values are
compatible with the present values of $\hat\sigma_{155}$ if the temperatures
in these samples were $T_{361901} \simeq 380^\circ$C and
$T_{521472} \simeq 450^\circ$C. [Actually, these temperatures disagree
with the Lutetium-derived ones: $T^{\rm Lu}_{361901} \simeq 250^\circ$C
and $T^{\rm Lu}_{521472} \simeq 280^\circ$C.  This difference is
probably to be explained by a moderate amount of contamination of natural
Lutetium after the reaction \cite{2.553}.] However, we could not use these data
to derive more stringent limits on $\alpha$ because the ${\rm Gd}^{155}$
resonance turns out to be less than half as sensitive as Samarium to changes in
$\alpha$ (see next Section).

\section{Translating possible resonance shifts into a bound on the variation
of the fine-struc\-ture constant.} \label{sec:4}

Let us finally translate the allowed range (\ref{eq:3.8}) into a bound on a
possible difference between the value of $\alpha$ during the Oklo
phenomenon and its value now. The treatments given in previous analyses are
unsatisfactory. The original analysis of Shlyakhter \cite{S76a,S76b,S83} rested
on a coarse representation of the nucleus as a square potential well, together
with dubious assumptions about nuclear compressibility, while the analysis of
Ref.~\cite{SV90} used an ill-motivated finite-temperature description of the
excited state of the compound nucleus.

The observed neutron-resonance energy $E^{\rm now}_{r} =0.0973$~eV, for
the radiative capture of neutrons by Sm$^{149}_{62}$, corresponds to the
existence of a particular excited quantum state of Sm$^{150}_{62}$.
More precisely, if we write the total mass-energy of the relevant
excited state of Sm$^{150}_{62}$ as \footnote{We set $c=1 $.}
$E^*_{150} = 62 m_p + 88m_n + E_1$ (with $E_1<0$), and the total mass-energy
of the ground state of Sm$^{149}_{62}$ as $E_{149} =62 m_p +87m_n +E_2$
(with $E_2 <0$), we have

\begin{equation}
 E_r = E^*_{150} - E_{149} - m_n = E_1 - E_2\ . \label{eq:1}
\end{equation}
Both $E_1$ and $E_2$ are eigenvalues of the Hamiltonian
\begin{equation}
   H = H_n + H_c\ ,  \label{eq:2}
\end{equation}
where $H_c$ is the Coulomb energy 
\begin{equation}
   H_c = e^2 \sum R_{ij}^{-1}\ ,   \label{eq:3}
\end{equation}
summed over the pairs of protons in the nucleus, and $H_n$ is, to a good
accuracy, independent of $e^2$.  We neglect small effects such as the
magnetic-moment interactions or the QED corrections to the masses, such
as $m_n$ and $m_p$, entering the nuclear Hamiltonian $H_n$.

Now let $e^2$ vary while $H_n$ remains fixed. Then for any eigenstate of
$H$ with eigenvalue $E$,
\begin{equation}
 {e^2}dE/d{e^2} = \langle H_c \rangle\ ,  \label{eq:4}
\end{equation}
and therefore
\begin{equation}
{e^2}d E_r/d e^2 = \langle H_c \rangle_1 - \langle H_c \rangle_2\ .\label{eq:5}
\end{equation}
The Coulomb energies on the right of (\ref{eq:5}) are not directly measurable.
The quantities that can be directly measured by optical spectroscopy
\cite{Otten} are the mean-square radii $\langle r^2 \rangle$ of the
charge-distributions of the protons in the various isotopes of
Samarium. Let us recall that ``isotope shifts'' in heavy atoms are related
to the effect of the finite extension of the nucleus on electron energies. A
first-order perturbation analysis of the latter effect (see, e.g.,
\cite{LL}) yields $\Delta E =(2\pi/3) \psi^2_e (0) Z e^2\langle
r^2\rangle$ where $\psi_e (r)$ is an (s-state) electron wave function, and
where $\langle r^2\rangle =Z^{-1} \int \rho r^2 dv$ with $\rho$ denoting
the proton charge distribution in the nucleus.

 To connect the expectation-values in (\ref{eq:5}) with the
mean-square radii, we use the semi-classical approximation
\begin{equation}
 \langle H_c \rangle_i = {1\over 2}  e^2\int V_i \rho_idv,
    \;i = 1,2\ ,\label{eq:6}
\end{equation}
where $\rho_i$ is the density of protons in the nuclear state $i$,
normalized to 
\begin{equation}
  \int \rho_i dv = Z = 62\ ,   \label{eq:7}
\end{equation}
for Samarium, and $V_i$ is the electrostatic potential generated by 
$\rho_i$. From (\ref{eq:5}) and (\ref{eq:6}),
\begin{equation}
 dE_r/d{e^2} = {1\over 2} \int({V_1}{\rho_1}
 -{V_2}{\rho_2})dv = -\left( {1\over 2} \right)\int{\delta}V\delta{\rho}dv
  + \int{V_1}\delta{\rho}dv < \int{V_1}\delta{\rho}dv\ ,  \label{eq:8}
\end{equation}
with
\begin{equation}
 \delta\rho = \rho_1 - \rho_2, \;\;\;  {\delta}V = V_1 - V_2\ . \label{eq:9}
\end{equation}

The term that is dropped in (\ref{eq:8}) is negative because it is minus an
electrostatic self-energy.  The integrand on the right side of (\ref{eq:8}) is
a small difference $\delta\rho$ multiplied by the smooth potential
$V_1$.  With an error of second order in small quantities, we may
approximate $V_1$ by the classical potential of a uniformly charged
sphere with radius $R_1$,
\begin{equation}
V_1(r) = Z \left[ {3R_1^2 - r^2\over 2R_1^3} \right]\ .  \label{eq:10}
\end{equation}
Then (\ref{eq:8}) becomes
\begin{equation}
 {d E_r\over d{e^2}} < -\left[ {Z^2\over 2R_1^3} \right]
       \delta_{12}(r^2)\ ,  \label{eq:11}
\end{equation}
where 
\begin{equation}
 \delta_{12}(r^2) = {1\over Z} \int{r^2}\delta{\rho}dv
        \label{eq:12}
\end{equation}
is the difference in mean-square charge-radius between the states $1$
and $2$.  

    Let the label $3$ denote the ground-state of ${\rm Sm}^{150}$.  Then
\begin{equation}
 \delta_{12}(r^2) = \delta_{13}(r^2) + \delta_{32}(r^2)\ .  \label{eq:13}
\end{equation}
The difference $\delta_{13}(r^2)$ cannot be calculated because we do
not know the shape of the excited state of ${\rm Sm}^{150}$.  But it
seems safe to assume that the proton charge distribution will not be
more tightly concentrated in the excited state than in the ground
state.  That is to say,
\begin{equation}
  \delta_{13}(r^2) \ge 0\ .  \label{eq:14}
\end{equation}

An inequality stronger than (\ref{eq:14}) could be deduced from more dubious
assumptions about nuclear compressibility, but a stronger inequality
is not needed.   From (\ref{eq:11}) and (\ref{eq:14}) we have
\begin{equation}
 {d E_r\over d{e^2}} < -\left[ {Z^2\over 2R_1^3} \right]\delta_{32}(r^2)\ ,
     \label{eq:15}
\end{equation}
and this is sufficient for our purposes.  The experimental isotope-shift
measurements reported by \cite{BSS} give directly $\delta_{34} (r^2) = 0.303
\pm 0.016 \, {\rm fm}^2$ and $\delta_{24} (r^2) = 0.092 \pm 0.005 \, {\rm fm}^2$,
where the label 4 denotes the ground-state of  ${\rm Sm}^{148}$. Taking the
difference gives 
\begin{equation}
  \delta_{32}(r^2) = 0.211 \pm 0.017 \;{\rm fm}^2 \ (3\sigma \ \hbox{error}) \ . 
\label{eq:16} \end{equation}
For the radius of the ${\rm Sm}^{150}$ nucleus to insert in (15), we
use equations (50) and (51) on page 568 of \cite{Otten}, which give 
\begin{equation}
  R_1 = 8.11 \;{\rm fm}.  \label{eq:17}
\end{equation}
{} From (\ref{eq:15}), (\ref{eq:16}) and (\ref{eq:17}), we find
\begin{equation}
 \alpha {dE_r\over d\alpha} < -(1.09 \pm 0.09)\; {\rm MeV}\ .  \label{eq:18}
\end{equation}

The estimate (\ref{eq:18}), obtained here
directly from measurements of the small charge-radius difference
(\ref{eq:16}) between Sm$^{150}$ and Sm$^{149}$, agrees with the result
obtained by differencing the phenomenological Bethe-Weizs\"acker formula
(droplet model). The latter formula estimates the nuclear Coulomb energy as
$\langle H_c\rangle =0.717\, Z(Z-1) A^{-1/3}$~MeV \cite{BW}. Taking
the difference between Sm$^{150}_{62}$ and Sm$^{149}_{62}$ (and arguing
as above that excited states are less charge concentrated) yields the
inequality $\alpha dE_r/d\alpha < -1.14$~MeV, which is compatible with the
result (\ref{eq:18}). By contrast, the Bethe-Weizs\"acker formula overestimates
by about a factor two the $\alpha$-sensitivity of the resonance energy
$E_r^{\rm now} = 0.0268 \, {\rm eV}$, for the radiative capture of neutrons by
${\rm Gd}_{64}^{155}$. This follows (using the same method as above) from the
fact that isotope-shift measurements reported in \cite{borisov} yield
\begin{equation}
\langle r^2 \rangle_{156} - \langle r^2 \rangle_{155} = 0.097 \pm 0.005 \, {\rm
fm}^2 \, , \label{eq:xx}
\end{equation}
which is less than half the Samarium difference (\ref{eq:16}). As a consequence
the $\alpha$-sensitivity parameter $\vert \alpha \, dE_r / d\alpha \vert$ of the
${\rm Gd}^{155}$ resonance is less than half that of the ${\rm Sm}^{149}$
resonance.

We are now in position to convert the bound (\ref{eq:3.8}) obtained
above from our analysis of Oklo data into a bound on the variation of
$\alpha$. To be conservative we use the worst $3\sigma$ limit on the
$\alpha$-sensitivity of $E_r$ obtainable from (\ref{eq:18}), namely
\begin{equation}
\left| \alpha {dE_r\over d\alpha}\right| > (1.09 -0.09)\;{\rm MeV}\;
      = 1.0 \;{\rm MeV}   \label{eq:19}
\end{equation}
Combining
(\ref{eq:19}) with the bound (\ref{eq:3.8}) on the shift $\Delta =E_r
(\alpha^{\rm Oklo}) - E_r (\alpha^{\rm now}) = - \vert \alpha dE_r/d\alpha
\vert (\alpha^{\rm Oklo} -\alpha^{\rm now})/\alpha$ yields our final result 
\begin{equation}
-0.9 \times 10^{-7} < {\alpha^{\rm Oklo} -\alpha^{\rm now} \over
\alpha} <
  1.2 \times 10^{-7}\ ,
  \label{eq:20}
\end{equation}
which we consider as a 95\% C.L. limit.

Though we have been very conservative in our analysis, our result (\ref{eq:20})
confirms the main claim of Refs.~\cite{S76a,S76b,S83}, namely that Oklo
rare-earth data are extremely sensitive probes of a possible variation of
$\alpha$: the surprisingly good $\sim 10^{-7}$ bound comes mainly from the
$10^7$ amplification factor between the MeV level in $\alpha dE_r/d\alpha$
(which is physically clearly understood from the Bethe-Weizs\"acker formula)
and the 0.1~eV level of the value of the Samarium resonance (with respect
to the threshold).

Let us note also the consistency of the approximations we made: a change
$\delta\alpha /\alpha \sim 10^{-7}$ has a totally negligible effect in all
the quantities (such as $\gamma_n$ or $\Gamma$) depending at most polynomially
on $\alpha$, and its effect on the dimensionless ratio $\hat\sigma_{149}
/\hat\sigma_{143}$ is dominated by the Breit-Wigner denominator of
$\hat\sigma_{149}$, i.e. by the change $\partial E_r^{149}
/\partial\alpha \, \delta\alpha$. [The contribution coming from the
shift of neutron capture resonances on ${\rm Nd}^{143}$ is relatively
negligible because, from the approximate $Z(Z-1)A^{-4/3}$ dependence expected
from the Bethe-Weizs\"acker formula, $\delta E_r^{143} \, \alt \, \delta
E_r^{149}\sim 0.1~{\rm eV}$ which is small compared to the near-threshold
resonances in ${\rm Nd}^{143}$, specifically the one below the threshold at
$E_r^{143}\sim -6~{\rm eV}$.]

In deriving the bound (\ref{eq:20}), we have implicitly assumed that,
during the Oklo phenomenon, $\alpha$ took some fixed value $\alpha^{\rm
Oklo}$ (possibly different from $\alpha^{\rm now}$). The situation would
be more complicated if, at the time, $\alpha (t)$ were oscillating on a
time scale smaller than the duration of Oklo. As Sm$^{149}$ data depend
essentially on the value of $\hat\sigma_{149}$ at the end of the fission
reaction, we expect that the bound (\ref{eq:20}) restricts the
amplitude of the variation of $\alpha$ in many extended scenarios
comprising $\alpha$-oscillations.

Dividing (\ref{eq:20}) by
the age of the Oklo phenomenon, we can convert it into a bound on the
time derivative of $\alpha$ averaged over the time span separating us
from the end of the Oklo phenomenon. [In scenarios where $\alpha$ varies
on the Hubble time scale, this averaged time derivative is nearly equal
to the present time derivative.]
This conversion introduces a further uncertainty, because the age of
Oklo is not determined with precision. The geochronological studies
suggest an age around $1.8\times 10^9$yr \cite{Oklo1}, while several
studies based on nuclear decay time scales gave $\sim 10\%$ higher
values: for instance, $1.98\times 10^9$yr \cite{2.441}, $1.93\times
10^9$yr \cite{2.495}, and $2.05\times 10^9$yr \cite{2.513}. To remain
conservative in our bounds, we shall use the lower, geochronological
value. Dividing Eq.~(\ref{eq:20}) by $1.8\times 10^9$yr, we get the
following conservative (95\% C.L.) limit on the time derivative of
$\alpha$ averaged over the time since the Oklo reactor was running
\begin{equation}
-6.7 \times 10^{-17} {\rm yr}^{-1} < {\dot\alpha \over \alpha} < 5.0 \times
10^{-17} {\rm yr}^{-1}
  \ . \label{eq:21}
\end{equation}
This is weaker than Shlyakhter's estimates (which ranged between $ \pm 5\times
10^{-18}$yr$^{-1}$ \cite{S76b} and $\pm 10^{-17}$yr$^{-1}$ \cite{S76a,S83})
 but rests on a firmer experimental basis. On the other hand, this is
between two and three orders of magnitude stronger than the other
constraints on the variability of $\alpha$ (see the Introduction).
Thanks to recent advances in atomic clock technology, it is conceivable
(and desirable) that direct laboratory tests might soon compete with
the Oklo bound (\ref{eq:21}).

We have focussed in this paper on the time variation of the fine-structure
constant because one can estimate with some confidence the effect of a change
of $\alpha$ on resonance energies. It is more difficult to estimate the effect
of a change in the Fermi coupling constant $G_F$, or, better, in the
dimensionless quantity $\beta = G_F \, m_p^2 \, c / \hbar^3 \simeq 1.03 \times
10^{-5}$. The estimates of Ref. \cite{hw76} for the (Weinberg-Salam)
weak-interaction contribution to nuclear ground state energies yield $E_{\rm
weak}^{150} - E_{\rm weak}^{149} \simeq 5.6 \, {\rm eV}$. If one assumes that
this gives an approximate estimate of the difference involving the relevant
excited state of ${\rm Sm}^{150}$, and that there is no cancellation between the
effects of changes in $\alpha$ and $\beta$, one finds from Eq. (\ref{eq:3.8})
the approximate bound
\begin{equation}
\frac{\vert \beta^{\rm Oklo} - \beta^{\rm now} \vert}{\beta} < 0.02 \, ,
\label{eq:22}
\end{equation}
\begin{equation}
\left\vert \frac{\dot{\beta}}{\beta} \right\vert < 10^{-11} {\rm yr}^{-1} \, .
\label{eq:23}
\end{equation}
This bound is more stringent than the limit $\vert \dot{\beta} / \beta \vert <
10^{-10} {\rm yr}^{-1}$ obtained from the constancy of the $K^{40}$ decay rate
\cite{Dy72}, and is comparable to the limit derived from Big Bang
nucleosynthesis: $\vert \beta^{\rm BBN} - \beta^{\rm now} \vert / \beta < 0.06$
\cite{bbn}.

Deducing from Oklo data a limit on the time variation of the ``strength of the
nuclear interaction'' poses a greater challenge. First, one must notice that,
as remarked in Section 2, only dimensionless ratios of nuclear quantities, such
as $\sigma_{149} / \sigma_{143}$, can be extracted from Oklo data. Within the
QCD framework, one generally expects any such dimensionless ratio to become a
(truly constant) pure number in the ``chiral'' limit of massless quarks.

Time variation of such a dimensionless ratio is then linked (in QCD) with
possible changes in the subleading terms proportional to the mass ratios $m_q /
m_p$, where $m_q$ denotes the masses of the light quarks. However, the chiral
limit of nuclear binding energies is tricky because of non-analyticity 
effects in
$m_q$. The present chiral perturbation technology does not allow one to
estimate the dependence of nuclear quantities such as $E_r^{149}$ or
$\sigma_{149} / \sigma_{143}$ on $m_q / m_p$. One, however, anticipates that
Oklo data might provide a very stringent test (probably at better than the
$10^{-7}$ level) on the time variation of $m_q / m_p$. To separate
unambiguously the effects of variations in $\alpha$ and $m_q / m_p$, it would
be necessary to extract from Oklo data several independent measured quantities
(e.g. by analyzing in detail the effects of resonance shifts in ${\rm
Gd}^{155}$ and ${\rm Gd}^{157}$).

\bigskip
\noindent{\bf Acknowledgments}

\medskip
We thank A. Michaudon and B. Pichon for informative communications about nuclear
data and M. Ganguli and E. Hansen for helping us to obtain the
information relevant to the Oklo phenomenon. The work of T. Damour at the
Institute for Advanced Study was supported by the Monell Foundation.

\newpage

\begin{figure}[p]
\caption{
Variation of the effective neutron capture cross section of Sm$^{149}$,
$\sigma_{149}$, as a function of a possible shift $\Delta = E^{\rm Oklo}_r
-E^{\rm now}_r$ in the lowest resonance energy, for several values of the
neutron temperature $T$. $\sigma_{149}$, $\Delta$ and $T$ (labelling the
curves) are measured in kbarn, eV and degree Celsius, respectively. The two
horizontal lines represent a conservative range of values of $\sigma_{149}$
compatible with Oklo data.}
\end{figure}

\begin{table}
\caption{Effective neutron cross sections of Sm$^{149}$ computed for 15
Oklo samples using published data.}
\begin{tabular}{|l|c|l|}
\hline
Sample       & Reference       & $\hat\sigma_{149}$ (kbarn) \\
\hline
KN50-3548    & \cite{R76}      & 93 \\
SC36-1408/4  & \cite{2.441}    & 73 \\
SC36-1410/3  & \cite{2.441}    & 73 \\
SC36-1413/3  & \cite{2.441}    & 83 \\
SC36-1418    & \cite{2.441}    & 64 \\
SC39-1383    & \cite{1.357,2.433}   & 66 \\
SC39-1385    & \cite{1.357,2.433}   & 69 \\
SC39-1387    & \cite{1.357,2.433}   & 36 \\
SC39-1389    & \cite{1.357,2.433}   & 64 \\
SC39-1390    & \cite{1.357,2.433}   & 82 \\
SC39-1391    & \cite{1.357,2.433}   & 82 \\
SC39-1393    & \cite{1.357,2.433}   & 68 \\
SC35bis-2126 & \cite{1.357,2.433}   & 57 \\
SC35bis-2130 & \cite{1.357,2.433}   & 81 \\
SC35bis-2134 & \cite{1.357,2.433}   & 71 \\
SC52 1472    & \cite{2.553}         & 72 \\
\hline
\end{tabular}
\end{table}

\begin{table}
\caption{Oklo data corresponding to the extreme cross-section results of
Eq. (\ref{eq:2.11}). Notation as in Eqs. (\ref{eq:2.7}), (\ref{eq:2.8}).}
\begin{tabular}{|l|c|c|c|c|c|c|c|} 
\hline
Sample  &$N_5/N_8$ &$N_{144}$ &$N_{147}$ &$N_{148}$ &$N_{149}$
&$\tau (10^{21}{\rm n/cm}^2)$ &$\hat\sigma_{149}$(kbarn) \\
\hline
SC35bis-2126 &0.00568 &0.22 &53.86 &2.39 &0.44 &0.92 &57 \\
KN50-3548    &0.00465 &0.16 &52.63 &6.90 &0.19 &1.25 &93 \\
\hline
\end{tabular}
\end{table}

\end{document}